\documentstyle[twoside,fleqn,espcrc2,epsf]{article}

\input{epsf}

\begin{document}

\title{Ratchet Effects for Vortices in Superconductors with
Periodic Pinning Arrays} 
\author{C. Reichhardt and C.J. Olson Reichhardt
\address{Center for Nonlinear Studies and Theoretical Division,
Los Alamos National Laboratory, Los Alamos, NM 87545, USA}}


\begin{abstract}
Using numerical simulations  
we show that novel transport phenomena can occur for vortices moving 
in periodic pinning arrays when two external perpendicular ac drives 
are applied. In particular, we find a ratchet effect where the
vortices can have a net dc drift even in the absence of a dc drive.
This ratchet effect can occur for ac drives which create orbits that break
one or more reflection symmetries.   
\vspace{1pc}
\end{abstract}

\maketitle


Recently there has been considerable interest in using 
ratchet effects to control the motion of vortices 
in superconductors \cite{1,2,3,4}.
In a ratchet, a net dc flow can arise under the application
of a strictly ac drive \cite{5}. 
Typically, the symmetry breaking which can 
allow for this effect is produced by an asymmetric underlying substrate,
such as a saw-tooth potential. However, ratchet effects can occur
in systems with {\it symmetrical} substrates when some {\it other} 
form of symmetry 
breaking is introduced. 
One possible source of such a symmetry breaking 
is the applied ac drive itself. For example, it was recently shown that
a particle moving in a two-dimensional (2D) 
periodic substrate can exhibit a ratchet effect
when crossed ac drives are applied, where the ac drives cause the
particle to move in orbits that have broken reflection symmetries \cite{6}. 
In this paper, we show that for vortices
moving in a periodic pinning array at fields $B/B_{\phi} > 1.0$ (where
$B_{\phi}$ is the field at which each pinning site captures one vortex),
a series of novel dynamical phases which ratchet the vortices can arise
when two external perpendicular ac drives are applied.      
Our system can be realized in superconductors with periodic pinning arrays
small enough that only one vortex can be captured per site, so that the
additional vortices sit in the interstitial regions between the pinning
sites \cite{7,8,9,10}. 

We simulate a thin-film superconductor 
containing an $N\times N$ square pinning array 
with a lattice constant $a$.
The equation of motion for a vortex $i$ is given by 
\begin{equation}
{\bf f}_{i} = \frac{d {\bf r}_{i}}{dt} = {\bf f}^{vv} + 
{\bf f}_{i}^{vp} + {\bf f}_{AC}.
\end{equation}
The force from the other vortices is 
${\bf f}_{i}^{vv} = -\sum_{j\neq i}^{N_{v}}\nabla_i U_{v}(r)$.
The vortices interact logarithmically via $U_{v} = -\ln(r)$. 
The force from the pinning sites is ${\bf f}_{i}^{vp}$. The pinning
sites are modeled as parabolic traps with a range $r_{p}$,
where $r_{p}/a = 0.1$, and maximum pinning force $f_{p}$. 
For the results in this
work, all external drive forces are much smaller than the pinning forces,
so that vortices in the pinning sites remain immobile. 
The ac drive is applied in both the $x$ and $y$ directions:
\begin{equation}
{\bf f}_{AC} = f^{ac}_{x}(t){\bf {\hat x}} + 
f^{ac}_{y}(t){\bf {\hat y}}.
\end{equation}
For all the ac drives, in the absence of a substrate there is no
dc drift velocity and $<f_{AC}> = 0.0$.
In the initial configuration,
all the pinning sites are filled with one vortex each, 
and the additional vortices are placed
randomly in interstitial locations. 
We monitor the long time average vortex velocities. 

We first consider the case of a system with $B/B_{\phi} = 1.065$, so that 
the interstitial vortices are far apart and in general do not interact. 
We apply an ac drive of the form 
${\bf f}_{AC} = A(\sin(\omega_{B}t) + \sin^{2}(\omega_{A}t)){\bf {\hat x}} + 
B\cos(\omega_{B}t){\bf {\hat y}}$,
where $\omega_{B}/\omega_{A} = 1.25$. In this case a symmetry breaking
arises from the shape of the orbits. 
In Fig.~1 we show $<V_{y}>$ 

\begin{figure}
\center{
\epsfxsize=0.5\textwidth
\epsfbox{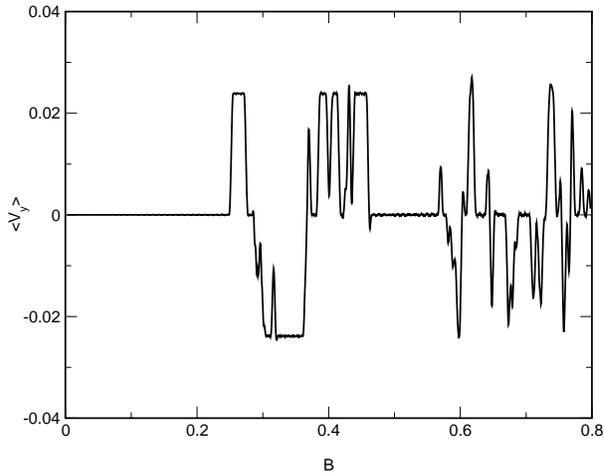}}
\caption{
The average velocity in the $y$-direction $<V_{y}>$ vs $B$, the 
coefficient of the $y$ component of the ac drive, for fixed $A = 0.49$.} 
\label{fig:fig1}
\end{figure}

\noindent
for a system
where $A = 0.49$ is fixed and $B$ is varied. 
For $B < 0.24$, the average vortex velocity is zero with the interstitial
vortices moving in closed orbits.  
For $B > 0.24$, there are a series of phases which have
a net dc flux in the $y$-direction. 
This flux can be in either the positive or negative direction.
In general most of the rectified phases give a drift velocity of 
$0.024$. We also find evidence for some rectifying phases that 
produce dc drifts that are fractional
multiples of the maximum drift value. 
However, these phases appear only for very small
regions of $B$. Additionally, as $B$ is further increased,
we find regions
where the interstitial vortices become repinned and 
$<V_{y}> = 0.0$. Each of the rectifying regions corresponds
to different dynamical phases where the vortices move in 
distinct periodic orbits.   

In Fig.~2 we illustrate some of the dynamical phases for
the system in Fig.~1. 
The positive rectifying phase at $B = 0.265$ is shown in Fig.~2(a).
Here a very intricate periodic pattern forms, with the vortices moving in 
a long time zig-zag pattern. In Fig.~2(b) the negative rectifying phase 
is shown for $B = 0.325$. Here another distinct 
orbit forms with
the vortex moving in lobes that slant alternately.
In Fig.~2(c) we show another positive rectifying mode 
at $B = 0.39$, where the orbit is similar to that in Fig.~1(b) but
the net drift is in the opposite direction. In Fig.~2(d) we illustrate
a non-rectifying orbit. Here the vortices move in complex periodic
closed orbits without a net drift. 

\begin{figure}
\center{
\epsfxsize=0.5\textwidth
\epsfbox{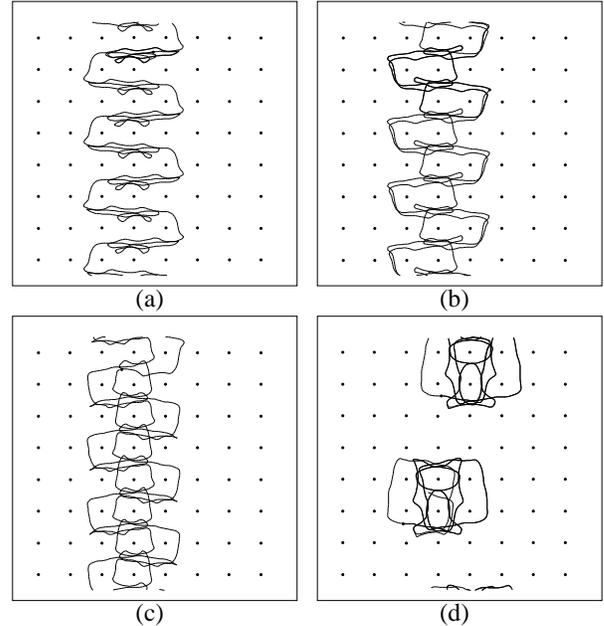}}
\caption{Vortex trajectories (black lines) and pinned vortex positions
(black dots) for the system shown in Fig.~1. (a) $B = 0.265$,
(b) $ B = 0.325$, (c) $B = 0.39$, and (d) $B = 0.5$.} 
\label{fig:fig2}
\end{figure}

We next consider the case where $B$ is fixed to $B=0.34$ while $A$ is varied. 
In Fig.~3 we plot $<V_{y}>$ vs $A$. Here we find similar behavior to
that in Fig.~1, with both positive and negative regions of 
net dc drift appearing along with pinned regions. In both Fig.~1 and 
Fig.~3, we find no ratchet effect for low values of $A$ or $B$.  

In Fig.~4 we illustrate some of the dynamical orbits for the system in Fig.~3.
In Fig.~4(a), we show the first rectifying phase which occurs at 

\begin{figure}
\center{
\epsfxsize=0.5\textwidth
\epsfbox{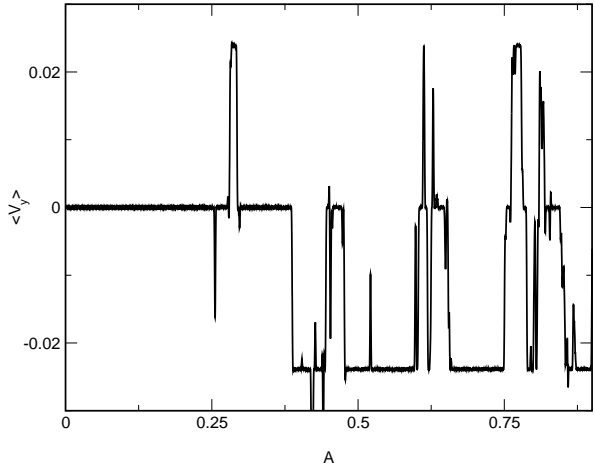}}
\caption{
The average velocity in the y direction $<V_{y}>$ vs $A$, the ac 
amplitude for the $x$ component of the
ac drive, at fixed $B = 0.34$.
}
\label{fig:fig3}
\end{figure}

\noindent
$A = 0.256$. Again, very intricate
periodic vortex motions occur. In Fig.~4(b) we plot the positive
rectifying phase that occurs at $A = 0.286$, where the 
net dc motion is in the positive $y$-direction. The orbit
consists of the vortex moving in loops around a pining site
with a series of much smaller sub-loops. In Fig.~4(c) we
show the pinned phase 
for $A = 0.289$ that occurs just after the phase in 
Fig.~4(b). In Fig.~4(d) we illustrate a rectifying phase at
$A = 0.792$. Here the orbit is much wider in  the $x$ direction, 
corresponding to the increased amplitude of the $x$ component of the ac drive. 
We find similar intricate orbits at the other rectifying phases, 
not shown here.

We have also considered other parameters and different ac drive
forms and find similar behaviors, 
indicating that the ratchet behaviors we observe are
a very general feature of 2D systems 
driven with perpendicular complex ac drives.

In conclusion, we have shown that with perpendicular ac drives
applied to  vortices in periodic pinning arrays,
a remarkably rich variety of dynamical phases can be achieved including
ratchet effects. 
The ratchet effect arises in this system 
even though the periodic
substrate is symmetrical due to a symmetry breaking by the 
ac drive itself.
Our results open a new avenue for controlling flux motion in superconductors.
For instance, it may be possible to exercise additional control over the
ratchet effect by changing the periodicity of the pinning array within 
a single sample. If, under the same ac drive, one type of pinning 
causes the vortices to move in the positive direction, 
while another type of pinning causes them to move in

\begin{figure}
\center{
\epsfxsize=0.5\textwidth
\epsfbox{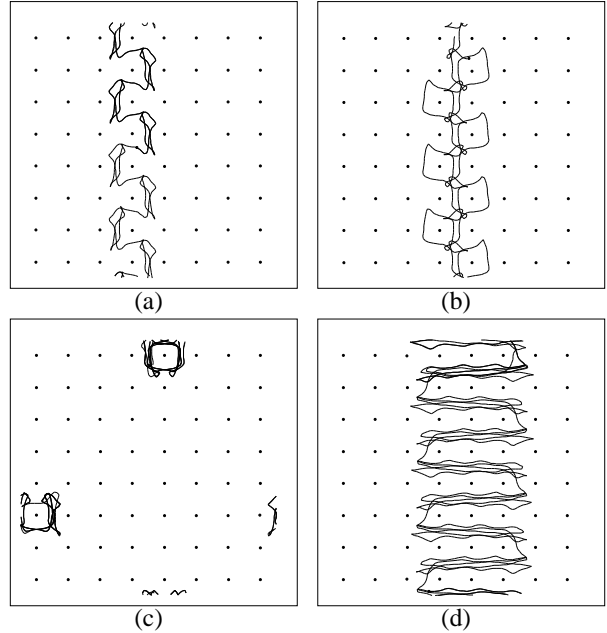}}
\caption{ The vortex trajectories (black lines) and 
pinned vortex positions (black dots) for the
system in Fig.~3 at: (a) $A = 0.256$, 
(b) $A = 0.286$, (c) $ A = 0.289$, (d) $A = 0.792$. 
}
\label{fig:fig4}
\end{figure}

\noindent
the negative direction, then it should be possible to create a fluxon
focusing effect where vortices can be constricted in small regions.
Conversely, it may be desirable to remove the flux from other
regions, which is useful for certain applications such as
flux sensitive devices. 

This work was supported by the US Department of Energy under Contract
No. W-7405-ENG-36.

\end{document}